\documentclass[a4paper,fleqn,usenatbib,useAMS]{mnras}
\usepackage{epsfig}
\usepackage{amsmath}
\usepackage{amssymb}
\usepackage{color}
\usepackage[T1]{fontenc}

\title{Night-sky brightness and extinction at Mt.~Shatdzhatmaz}

\author[V. Kornilov et al.]{V.\,Kornilov\thanks{E-mail: victor@sai.msu.ru}, M.\,Kornilov, O.\,Voziakova, N.\,Shatsky, B.\,Safonov  \newauthor I.\,Gorbunov, S.\,Potanin, D.\,Cheryasov, V.\,Senik \\ Lomonosov Moscow State University, Sternberg Astronomical Institute, Universitetsky pr-t, 13, 119234 Moscow, Russia.}

\begin{document}

\date{Accepted 2016 July 25.  Received 2016 June 26; in original form 2016 April 5}
\pagerange{\pageref{firstpage}--\pageref{lastpage}}
\pubyear{2016}

\maketitle
\label{firstpage}

\begin{abstract}
The photometric sky quality of Mt.~Shatdzhatmaz,  the site of Sternberg Astronomical Institute Caucasian Observatory 2.5~m telescope, is characterized here by the statistics of the night-time sky brightness and extinction. The data were obtained as a by-product of atmospheric optical turbulence measurements  with the MASS (Multi-Aperture Scintillation Sensor) device conducted in 2007--2013. The factors biasing night-sky brightness measurements are considered and a technique to reduce their impact on the statistics is proposed.

The single-band photometric estimations provided by MASS  are easy to transform to the standard photometric bands. The median moonless night-sky brightness is $22.1$, $21.1$, $20.3$, and $19.0$ mag per square arcsec for the $B$, $V$, $R$, and $I$ spectral bands, respectively. The median extinction coefficients for the same photometric bands are $0.28$, $0.17$, $0.13$, and $0.09$ mag. The best atmospheric transparency is observed in winter.
\end{abstract}

\begin{keywords}
techniques: photometric -- atmospheric effects -- site testing
\end{keywords}

\section{Introduction}

It is well known that the efficiency of classical ground-based astronomical observations (resolution and limiting magnitude) depends strongly on the atmospheric  seeing~\citep{Bowen1964}, but other astroclimatic parameters also play an important role. An overall telescope productivity is proportional to the clear sky fraction. Observations of faint object  depend strongly  on the night-sky brightness, which includes a contribution from the light pollution. The accuracy of photometric observations is determined by  temporal and spatial stability of the atmospheric extinction.

Since the Mt.~Shatdzhatmaz on North Caucasus had been chosen as a place for the new 2.5~m telescope, we initiated in 2007 a long-term monitoring of the atmospheric seeing and other astroclimatic parameters at this site, pursuing two main goals: first, support the operation of the 2.5~m telescope, and, second,  gain a better understanding  of the astroclimate of North Caucasus.

The first results of these measurements are presented in \citep{2010MNRAS} where the used hardware and technique are also described. The summary results of the optical turbulence studies of the 2007--2013 campaign  are given in the paper \citep{2014PASP}. In this paper we present the results of measurements of the photometric parameters, the night-sky brightness and the atmospheric extinction.

The night-sky background was measured during optical turbulence monitoring with the MASS device \citep{2007bMNRAS} because it is required to properly calculate scintillation indices \citep{2003MNRAS, 2007aMNRAS}. More than $30\,000$ such sky background estimates were obtained  during the campaign. 

Since MASS is essentially a fast high-precision photoelectric photometer, it also allows us to estimate atmospheric extinction.  To this end,  the optical turbulence program was complemented by special extinction measurements in 2009. 

The first results of these measurements are given by \citet{Voziakova2012AstL}. More comprehensive atmospheric extinction data are presented in Section~\ref{sec:extinction}. In the same Section, it is shown how to transform the extinction measured in the MASS spectral band to the standard $U$, $B$, $V$, $R$, and $I$ photometric system.

The statistics of sky background estimates are given in Section~\ref{sec:background}. They are transformed to the conventional units of stellar magnitude per square arcsec using the known magnitudes of the MASS program stars and the atmospheric extinction.

The results presented in this paper allow us to comprehensively characterise the Sternberg Astronomical Institute (SAI) observatory, facilitating the optimum scheduling of the 2.5~m telescope. Monitoring of the astroclimatic parameters  will help  to operate the  telescope in the most efficient way.

\section{Facilities description}

\label{sec:gen}

\subsection{Astroclimatic campaign of 2007--2013}

The Caucasian Observatory of SAI is located close to the Mt.~Shatdzhatmaz summit (Russia, North Caucasus,  Karachay-Cherkess Republic, about 20~km to the South from the city of Kislovodovsk). The mountain (the summit altitude is 2127~m above sea level) belongs to Skalisty ridge. The ridge is parallel to the Main Caucasian ridge that is 50~km to the South. The Skalisty ridge extends to the West and then turns to the North, forming the North Caucasus watershed divide.

The 2.5~m telescope is installed 40~m from the steep side  of the mount at the altitude of 2112~m above the sea level. The dome coordinates are $43^\circ 44^\prime 10^{\prime\prime}\mathrm{\,N}$, $42^\circ 40^\prime 03^{\prime\prime}\mathrm{\,E}$. Smaller telescopes are to be mounted in 5~m towers along the ridge, to the North-East from the main telescope. The facilities and the observatory infrastructure are located at a distance of 400~m from the towers, in a flat area. The solar Station of the Pulkovo observatory is located at 700~m to the North-West.

The southern part of the sky is free from the light sources, while the northern part is illuminated by the townships of the Caucasian Mineral Water region. The stratovolcano Mt.~Elbrus (altitude 5624~m) is located at a distance of 47~km South-South-West from the observatory.

There are other astronomical observatories in the region. The Special Astrophysical Observatory of Russian Academy of science is located 100~km to the West, at the northern spurs of the Main Caucasian ridge. The Terskol Astronomical Observatory is located on the slope of the Elbrus. Unlike the SAI observatory, both of them are located close to higher peaks.

The Automatic Site Monitor was installed in summer of 2007  on the top of  Mt.~Shatdzhatmaz, 40~m to the South-West from the 2.5~m telescope. Optical turbulence measurements have been carried out from October 2007 to June 2013. The monitor hardware and its operating principles have been described by us earlier \citep{2010MNRAS}; these publications focus on the optical turbulence measurements with the combined MASS/DIMM device.

The following characteristics for the atmosphere over  Mt.~Shatdzhatmaz have been established during the 2007--2013 campaign \citep{2014PASP}:

\begin{enumerate}
\item The median seeing $\beta_0$ is $0.96$~arcsec and it is better than $0.74$~arcsec in 25  per cent of time. The most probable value for the seeing is $0.81$~arcsec. The free atmosphere (above the 1~km) seeing is $0.43$~arcsec. The best seeing is observed in October--November, when the median seeing is $\approx 0.83$~arcsec. The optical turbulence is strongest in March ($1.34$~arcsec). 

\item The median isoplanatic angle is $2.07$~arcsec, also typical for many observatories. Its maximal value of $2.50$~arcsec is reached in October. The median atmospheric time constant $\tau_0$ is $6.57$~ms, increasing to  $10$~ms in autumn.

\item The clear sky time \citep{2016AstrL} is equal to 1320 hours per year or 45  per cent of the astronomical night time. The majority of the clear night time ($\approx 70$  per cent) is  concentrated in the period from the middle of September to the middle of March. The maximum fraction of clear sky amounts to $\approx 60$  per cent in November.
\end{enumerate}

\subsection{Photometric properties of the MASS instrument  }

The combined MASS/DIMM device uses both amplitude and phase distortions of the light coming through the atmosphere from the point source to measure atmospheric optical turbulence. The instrument  and methods are described in detail in the paper \citep{2007bMNRAS}. Below we briefly recall some important issues relevant to this paper.

The MASS/DIMM instrument is designed for usage with common amateur 25--30~cm telescopes,  convenient  in field astroclimatic campaigns. The light coming from the telescope is directed to the different channels of the instrument by the specially designed optical beam-splitter (segmentator). The segmentator divides the input pupil into different subapertures. The MASS channel has four similar detectors corresponding to the four input coaxial apertures: A, B, C, and D. Their diameters range from $2$ to $9$~cm.

For photometry, the apertures areas ($3$, $6$, $20$ and $30\mbox{ cm}^2$ for the A, B, C, and D channels, respectively) are essential. Photomultipliers R7400P with stable fast electronics are used to measure the light fluxes in photon counting mode with a dead time of $\approx 20$~ns. The dark count of these bialkali photomultipliers is only a few pulses per second under typical night-time ambient temperatures. In comparison, the mean dark night-sky background signal is about 200~pulses per second in the channels C and D. The measurements are carried out with microexposures of 0.5--1~ms, then the statistical moments are calculated for the counts.

The spectral response of MASS is determined both by the detector response curve and by the optics transmission\footnote{\url{http://curl.sai.msu.ru/mass/download/doc/} \url{mass_spectral_band_eng.pdf}}. The latter is especially important in the blue part of spectrum. The effective wavelength of the resulting photometric response ranges from $\sim 450$~nm to $\sim 500$~nm for different copies of MASS being used at various observatories over the world. The MASS spectral response should be thoroughly studied because it is required for the correct measurement of the turbulence vertical profile. 

Given non-linearity of the photon counters is properly accounted for \citep{2008ARep,2014JOSA}, the MASS device allows  precise photometry in the MASS photometric band for stars of $0$--$7$ magnitude. The scintillation noise is the primary source of photometric errors for these bright stars. Because of this noise,  the minimum exposure time needed to reach the precision of $0.001$~mag can be as long as $100$~sec  \citep{2012AA}.

\section{Atmospheric extinction}
\label{sec:extinction}

\subsection{Observations}

The optical turbulence measurement program was modified in January 2009 by including special observations to fulfill two tasks. Equal-altitude stars were measured to obtain the stellar magnitudes of program stars  outside atmosphere, to produce a catalogue of photometric standards  in the MASS band. Different-altitude stars were observed to measure the atmospheric extinction coefficient by means of the classical photometric pair technique.

The first report on the atmospheric extinction above the observatory is given by \citet{Voziakova2012AstL}. Since then, the amount of photometric observations has grown considerably. The number of  measurements of low-altitude standards (at air masses $M(z)$ of $1.35$--$2.0$) has  grown to $\approx 2500$. The number of observed equal-altitude pairs has also increased, the magnitudes of the  photometric standards have been refined, and their list has been extended to 33 objects.

Measurements of photometric standards were carried out every 1.5~hours if atmospheric transparency was stable within $\sim 0.02$ mag over 10--15 minutes. Such conditions were interpreted as a photometric sky condition. The set of  photometric measurements consists of  2 or 3 stars: an equal-altitude pair, an extinction pair, or both. One photometric measurement cycle takes 10--15 minutes.

Usually, the sky background is measured for 10~seconds after each telescope pointing. The photometric standard measurement is carried out for two minutes in the same manner as the optical turbulence  measurement. The precision for the flux estimation is limited by the scintillation noise, that is $\sim 0.01$~mag at zenith in the C and D MASS channels for 1 second exposure time. For a two minutes exposure, the precision is about $0.001$ mag. It may degrade to $0.005$ mag for low objects. The contribution of the photon noise is less than $2\cdot10^{-4}$ for any used star.

The pair technique provides reliable results only under photometric conditions because it is based on the assumption  that  the extinction coefficient $\alpha$ is the same in the directions to both stars. The conditions are called photometric when the sky is absolutely clear and uniform. Moreover, the  stars in the pair are required to have a substantial air mass difference $\Delta M$. An instrumental constant $R$ is expressed using the extinction coefficient $\alpha$ as follows:
\begin{equation}
R = m_* + \alpha\,M\left(z\right) + 1.086\,\ln F_*\left(z\right),
\label{eq:instr}
\end{equation}
where $F_*\left(z\right)$ is the measured star flux and $m_*$ is the magnitude of the photometric standard.

The instrumental constant connects the entrance pupil illumination to the measured signal. It depends on the detector sensitivity and the  transmission of the optics. Variations of the instrumental constant are caused by the MASS detectors behaviour, the ambient temperature, the optics contamination, the optics misalignment. Since all these factors do not change the quantity $R$ between two sequential measurements, they do not affect the estimate of the extinction coefficient $\alpha$. Here we use the counts in the C channel, considering that the D channel is more sensitive to misalignment and fogging.

To analyse the atmospheric extinction, we use the measurements obtained with the MASS/DIMM device number MD09. The device MD41 installed in January 2013 has a substantially different spectral response. The difference between these devices for   ``red'' stars ($B-V \approx 1.5$~mag) is up to $0.2$ magnitude.

\subsection{Extinction coefficient in the MASS photometric band  }

Uncertainties for the catalogue magnitudes $m_*$ of the photometric standards may lead to  errors of the extinction coefficient. In our case, the magnitudes are known with an accuracy of better than $0.01$ mag \citep{Voziakova2012AstL}. The effects of the photometric band width  (the colour term) can be neglected because only ``white'' stars ($B-V < 0.4$~mag) are used to measure the extinction coefficient. These errors, including the flux measurement error, are magnified by $1/\Delta M$ factor when the extinction coefficient $\alpha$ is calculated. 

We determined the minimum acceptable difference $\Delta M$ between the extinction stars by comparing the distributions for the extinction coefficient $\alpha$ as a function of this threshold. The shape of the  distribution is almost constant as long as $\Delta M > 0.2$. Because the sample size decreases as the airmass threshold increases, the final threshold was chosen to be $\Delta M > 0.25$.  Given this threshold, $1786$ extinction coefficient estimates are obtained.

The extinction coefficient $\alpha$ may be verified using the instrumental constant $R$. We assume that the instrumental constant varies smoothly over the time and its two consecutive measures should be similar. However, random variations of  $R$ are introduced by the fluctuations of ambient temperature, uncertainties of the extinction coefficient, and other measurement errors. The distribution of such random variations of $R$ has a Gaussian shape with the standard deviation of $0.025$~mag and wide tails extending to $\pm 0.1$~mag.

\begin{figure}
\centering
\psfig{figure=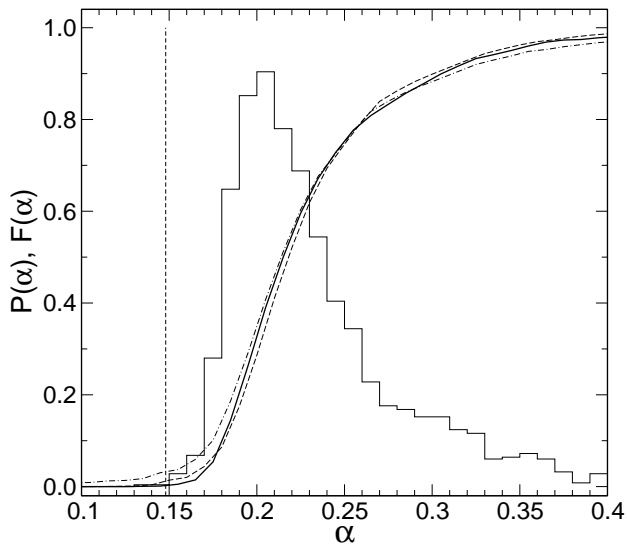,width=7.60cm,angle=-90}
\caption{The probability density function of the final estimates of the extinction coefficient $\alpha$ in the MASS spectral band is plotted by the  thin stepped line. The corresponding cumulative probability function is shown by the thick line. The dashed line shows the distribution for the initial estimations. The stroked line is  the distribution after filtering $7$ per cent of the data. The vertical line denotes the extinction coefficient for a pure Rayleigh atmosphere. \label{fig:alpha_distr} }
\end{figure}

The criterion for filtering outliers of the instrumental constant and the corresponding coefficient $\alpha$ estimates is formulated by requiring that any quantity $R$ must be close to the median of its 11 nearest neighbours, to within $\pm 0.1$~mag. About $7$ per cent of all estimates, including obvious outliers, have been excluded by this procedure. The cumulative distribution of the filtered extinction coefficients $\alpha$ is shown in Fig.~\ref{fig:alpha_distr} by the dashed line.

On the other hand, one can estimate the extinction coefficient using the expression  $\alpha\,M\left(z\right) = R - m_* - 1.086\,\ln F_*$, based on the near-zenith star flux and the smoothed instrumental constant. The distribution for $\approx 11\,000$ samples obtained by this alternative formula is in agreement with the previous one. Such approach is  useful mostly for real-time monitoring of atmospheric transparency.

A combined, robust estimate of the extinction coefficient  $\alpha$ can be obtained by averaging the results of the pair method and those of the alternative method based on the smoothed instrumental constant. It  should be more reliable because it uses more information. Probably, a detailed error analysis would allow us to determine optimal weighs of such averaging, but here we use a purely empirical approach. The resulting distribution of the average $\alpha$ estimates becomes a little narrower than the initial one.

The probability density function and the corresponding cumulative distribution of $\alpha$ are shown in Fig.~\ref{fig:alpha_distr}. One may see that the quantiles of the distribution are in good agreement, to within $0.01$~mag, independently of the filtering approach. The most probably value is $\alpha = 0.20$~magnitude. In the figure, Rayleigh atmosphere extinction coefficient calculated for the summit altitude by {\tt libRadtran} (see Sect~\ref{sec:stand_phot}) is plotted as a reference. The final distribution quantiles are given in Table~\ref{tab:alpha_quant}.

\begin{figure}
\centering
\psfig{figure=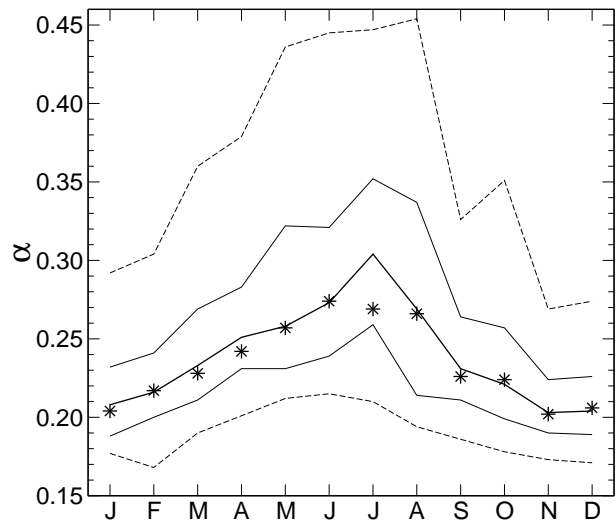,width=7.60cm,angle=-90}
\caption{
Variations of the quantiles of the extinction coefficient  
$\alpha$  over the year. The medians  of the extinction measured by the pairs method  are plotted by the $*$  symbols.  The thick solid line shows the medians for the $R$-based extinction estimates. The dashed lines show the 5 and 95  per cent marginal quantiles. The thin lines show the 25 and 75  per cent quantiles. \label{fig:alpha_monthly} }
\end{figure}

The annual variation of the extinction is shown in Fig.~\ref{fig:alpha_monthly}. The number of extinction measurements by the pair method is $\lesssim 100$ during spring and summer months. This makes it impossible to reliably determine the marginal quantiles. 

For this reason, for any near-zenith star measurement we use the extinction estimated from the smoothed instrumental constant. The quantiles for this sample are show in the Figure. The median curves are in good agreement for all months except July. However, there is no doubt that the best transparency is observed in November--January and the worst one is observed in June--August. The probability of a night with poor transparency ($\alpha > 0.4$~mag) is considerably higher in summer months.

\subsection{Extinction in the standard photometric system }
\label{sec:stand_phot}

\begin{table}
\caption{The quantiles of the extinction coefficient $\alpha$ for the MASS band and for the standard photometric bands, in magnitudes. \label{tab:alpha_quant}}
\centering
\begin{tabular}{l|rrrrr}
\hline
Photometric band & 5\% & 25 \% & 50 \% & 75 \% & 95 \% \\
\hline
MASS        & 0.17 & 0.19 & 0.22 & 0.25 & 0.34 \\
U           & 0.46 & 0.49 & 0.51 & 0.55 & 0.67 \\
B           & 0.23 & 0.25 & 0.28 & 0.31 & 0.42 \\
V           & 0.13 & 0.15 & 0.17 & 0.20 & 0.28 \\
R           & 0.10 & 0.11 & 0.13 & 0.16 & 0.23 \\
I           & 0.06 & 0.08 & 0.09 & 0.11 & 0.17 \\
\hline
\end{tabular}
\end{table}

The MASS photometric band is located between the standard $V$ and $B$ bands, closer to the latter. For our device MD09, the  effective wavelength is 479~nm for stars of spectral class A0\,V. This spectral range is situated in the convenient part of the atmospheric transmission window where the extinction is determined only by the Rayleigh and aerosol scattering. Other atmospheric components (ozone and water) scarcely affect the extinction in the MASS band. However, their impact may be considerable in other photometric bands.

We use the package {\tt libRadtran} \citep{Mayer2005ACP} to calculate the atmospheric transmission in the spectral range of 300--1200~nm. The atmospheric  structure is defined by the standard mid-latitude atmosphere model from the paper \citep{Shettle1990}. The aerosols are described by the model number 1 from that paper. Their concentration is given by the specific input parameter ``visibility'' measured in kilometres in {\tt libRadtran}. The precipitable water vapour is specified as usual in mm. The ozone concentration is specified in the Dobson units ($100\,DU = 1$~mm).

\begin{figure}
\centering
\psfig{figure=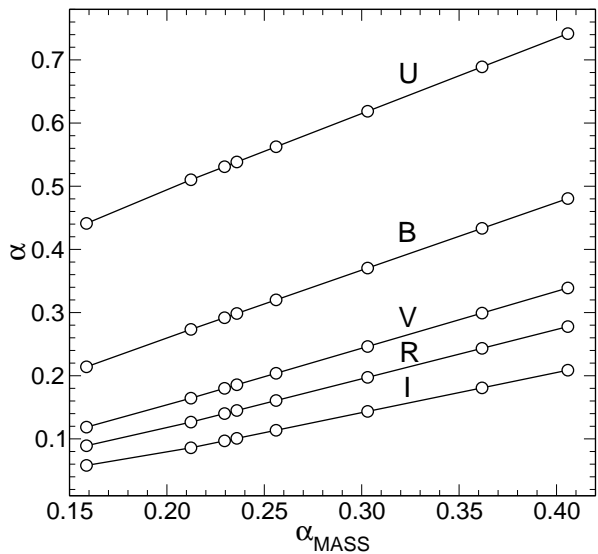,width=7.60cm,angle=-90}
\caption{Calculated dependence of the extinction coefficients $\alpha$ (circles) for photometric bands $U$, $B$, $V$, $R$, $I$ on the measured extinction coefficient for the MASS photometric band. \label{fig:ubvri_connect} }
\end{figure}

The aerosol concentration obviously makes a major impact on the atmospheric extinction variance. The visibility parameter variation range for the simulation was chosen to cover the whole observed range of the  MASS-band extinction. The precipitable water vapour amount was set to the median value $7.65$~mm corresponding to the results of  \citet{Voziakova2012AstL}. The ozone concentration was determined by the MSR data of the project ``Monitoring atmospheric composition \& climate'' \citep{Allaart2015AMT}~\footnote{\url{http://www.temis.nl/macc/index.php?link=o3\_msr\_intro.html}}. The ozone concentration varies from $280$ to $400$~DU, its median is $315$~DU.

The spectral transmission curves obtained by the simulation with {\tt libRadtran} were convolved with the source spectrum of $\alpha$~Lyr and the response curves for the bands $U$, $B$, $V$, $R$, $I$, and MASS. The results are expressed as extinction coefficient $\alpha$, as a function of the visibility parameter. Alternatively, they can be expressed as a function of the MASS-band extinction. The dependencies are shown in Fig.~\ref{fig:ubvri_connect}. Since they are almost linear,  the extinction measured with the MASS instrument can be easily transformed to the extinction coefficient for any standard photometric band. Small deviation from linearity arises because of the band width effects. It should be stressed that the dependencies are calculated for the specific MD09 device spectral response.

The extinction coefficient quantiles obtained by the described technique are given in Table~\ref{tab:alpha_quant}. The transformation stability has been proven by varying ozone and water vapour concentrations within their observed ranges. The precipitable water vapour was varied from $1$ to $20$~mm \citep[see][]{2016AstrL}, the ozone was varied from $280$ to $400$~DU. The impact of ozone variations on the extinction coefficient is less than $0.005$~mag. The precipitable water vapour variation affects mostly the bands $R$ and $I$, where its impact attains $0.008$ and $0.012$~mag, respectively.

\section{Night-sky brightness}
\label{sec:background}

\subsection{Specifics of the sky brightness measurement with MASS }

Let us recall that when the sky brightness is measured with photo multipliers, the contribution of faint stars is a major problem  \citep{Krisciunas1997PASP, Leinert1995AA, Mattila1996AA}. It follows from the fact that a large field aperture (1 -- 2 arcmin) has to be used in order to obtain statistically significant results. The only way to minimise the contamination was to choose a star-free sky patch intentionally. The MASS field aperture is even larger, its angular diameter is about $4$ arcmin and in our observations the fields were chosen randomly in the vicinity of bright program stars because the task of night-sky brightness measurement  was addressed only {\em a posteriori}.

Most observations have been carried out with the MD09 device; its aperture area is $39\,800$~square arcsec ($0.003$~square degrees). Given that the night-sky brightness is expected to be close to $22$ mag$\cdot$arcsec${}^{-2}$, the corresponding sky signal is equivalent to a star of $10.5$~mag when such a large aperture is used. As follows from the probabilistic estimation based on the star density \citep{Allen1973}, the mean overall star brightness in the MASS aperture is close to $\approx 11.4$~mag and  corresponds to an addition of $\sim 0.4$~mag to the sky signal.

Scattered light of the bright program stars is another effect  leading to an overestimation of the sky background. The typical sky signal is $\sim 10^4$ times fainter than the program star. The telescope mount randomly moves by $5$ arcmin to measure the sky background, but it appears that contribution of the scattered light  remains considerable at such angular distance. Actually, our preliminary analysis  has put in evidence the  statistical correlation between the sky brightness and the magnitude of the program star.

\begin{figure}
\centering
\psfig{figure=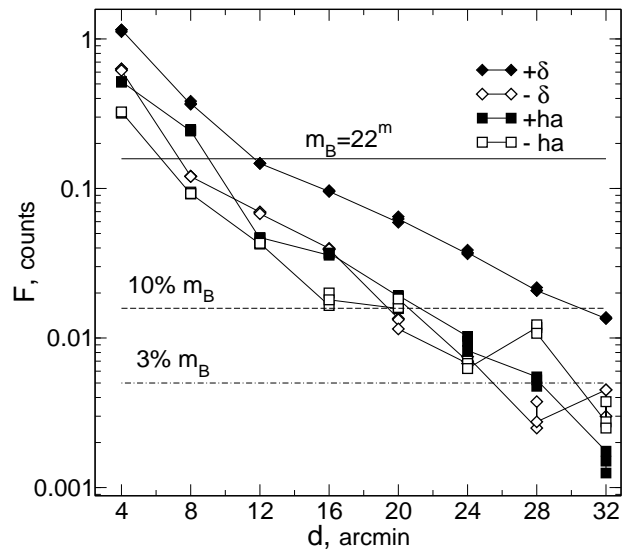,width=7.60cm,angle=-90}
\caption{Scattered light (sum of the counts in the channels C and D), normalised to $\alpha$~Lyr stellar magnitude, versus the distance from Sirius for four directions (see legend in the plot). The horizontal lines denote the dark night-sky brightness, and 10 and 3 per cent of it. \label{fig:scatter} }
\end{figure}

Dedicated measurements of the background around Sirius (MASS magnitude is $-1.5$) were carried out in January 2016. They show that the scattered light for $\alpha$~Lyr, the brightest program star, is about $\sim 10$  per cent of the signal from a dark sky at the distance of $20$ arcmin. The scattered light depends not only on the magnitude, but also on the direction of the angular offset; it is larger for offsets towards  the instrument's viewer (see $+\delta$ in Fig.~\ref{fig:scatter}). Indeed, the light of the star displaced from the optical axis by $30$~arcmin is not fully blocked because of the telescope and instrument geometry.

As we pointed out before, the sky background have been measured during the whole campaign, because it is required for the optical turbulence measurement. However, only the data obtained since the start of regular observations in March 2008 are used here to estimate the astronomical night-time sky brightness. After rejecting  the outlying points, there are $17\,162$ measurements in total for any MASS channel, each mostly 10 seconds long or at least 9 seconds long.

\subsection{Star light and scattered light effects}

Recording of the background in MASS is made in the same way as the stellar scintillation measurement, and hence it contains the spatio-temporal correlation of the counts with the 1~ms resolution. The basic assumption for the following method of  star light estimation is that there are no atmospheric scintillation and no temporally correlated signal in the sky background flux.

Let us denote the temporal covariance of the flux $F$  with a 1~ms lag by $\mathsf{Cov}_1[F]$. This statistic is strictly to zero for the sky background and close to the scintillation variance for the star flux. The covariance  appears to be the most suitable diagnostic of the scattered light because, unlike the variance, it does not need  precise evaluation of the Poisson noise contribution. 

For the real data, the sample covariance $c_1$ is calculated instead of $\mathsf{Cov}_1[F]$. Therefore, the statistic $C_1$, the ratio of $c_1$ to the sample mean, has a zero expectation and the following variance:
\begin{equation}
\mathsf{Var}[C_1] = \mathsf{Var}\left[\frac{c_1}{\bar F}\right] = \frac{1}{N},
\label{eq:7}
\end{equation}
where $N$ is the number of  flux counts. For the exposure time of 10~second, $N \approx 10\,000$. The flux $F$ is  the number of registered pulses per 1~ms (counts).

When a correlated stellar signal $F_*$ is present together with the background count $F_\mathrm{B}$, the statistic $C_1$ increases and its expectation becomes
\begin{equation}
\mathsf{E}[C_1] = \frac{F_*^2\,s_1^2}{F_*+F_\mathrm{B}},
\label{eq:8}
\end{equation}
where the covariance index $s_1^2$ is defined by the input aperture geometry and the atmospheric optical turbulence. In average, it is $\sim 0.02$ for the C and D channels. It is clear that we can increase the effect by summing $C_1$ for two or more channels, using the fact that the flux ratio of any two  channels is a constant value.

\begin{figure}
\centering
\psfig{figure=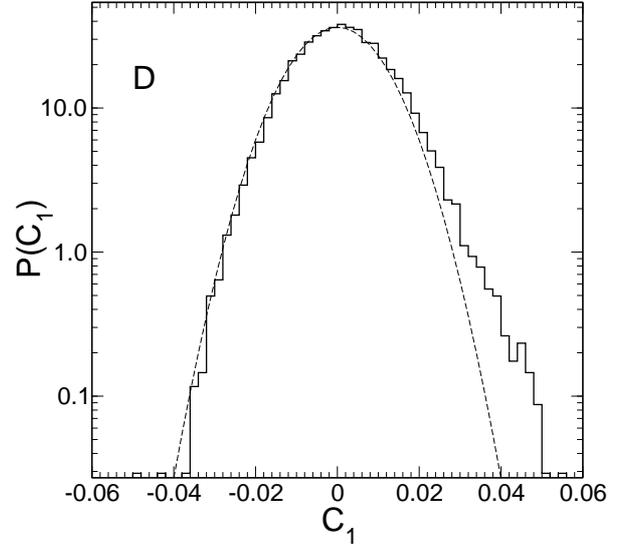,width=7.60cm,angle=-90}
\caption{The stepwise curve is the distribution of $C_1$ for the overall set of $17\,162$ measurements in the D channel. The dashed line is the theoretical Gaussian distribution according to Eq.~(\ref{eq:7}).
\label{fig:distD} }
\end{figure}

Fig.~\ref{fig:distD} displays the empirical distribution of the statistic $C_1$ and its approximating normal distribution with a variance $\sigma^2$ calculated using the expression (\ref{eq:7}). One can see the excess of positive values. Unfortunately, for our measurement parameters, the events with $C_1 > 3\,\sigma$ correspond to relatively bright  stars that exceed the sky flux and occasionally ($\sim 3$  per cent) appeared in the device aperture. Because of that, we statistically account for the stellar ``pollution'' further. This correction is valid 
only statistically, rather than for the individual estimations.

\begin{figure}
\centering
\psfig{figure=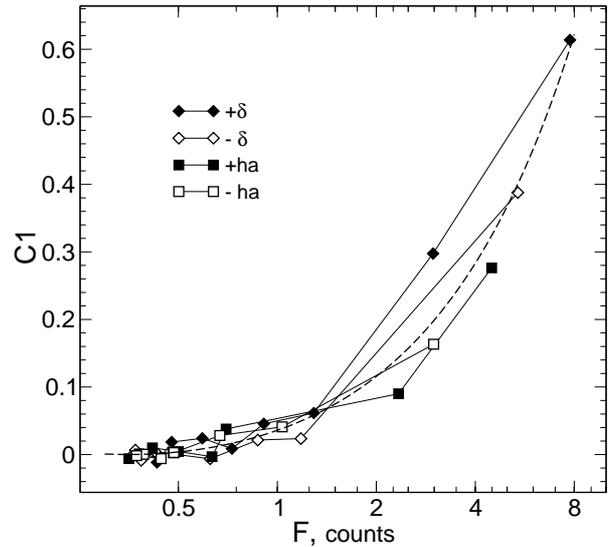,width=7.60cm,angle=-90}
\caption{Dependence of the statistic $C_1$ on the flux in the D channel for scattered light from Sirius,  in four offset directions. The dashed line denotes the approximation using Eq.~(\ref{eq:8}). \label{fig:scatter_C1} }
\end{figure}

The  dependence of $C_1$ on the total signal in field aperture (the sky background and the scattered light) obtained during measurement of the scattered light from Sirius is given in Fig.~\ref{fig:scatter_C1}. This dependence agrees well with Eq.~(\ref{eq:8}) whose prediction is shown by the dashed line for $s_1^2 \approx 0.08$ corresponding to the covariance observed at the moment of the experiment. This way, we assume that the scattered light ``pollution'' is statistically equivalent to the faint star contribution. Below both effects are  considered jointly.

\begin{figure}
\centering
\psfig{figure=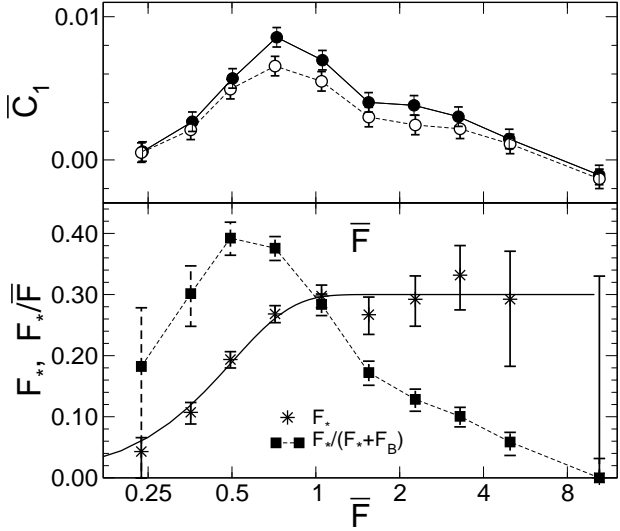,width=7.60cm,angle=-90}
\caption{Top panel: 10 subsequent sub-sample medians for the full sample of $17\,162$ background estimates ($\bullet$) and  after filtering the $\pm3\sigma$ outliers ($\circ$). Bottom panel: the stellar flux $F_*$ in counts ($*$) and its impact on the total mean flux $\bar F = F_*+F_\mathrm{B}$  ($\scriptscriptstyle\blacksquare$) in the D aperture. The solid line shows the $F_*$ approximation by Eq.~(\ref{eq:appro}).  \label{fig:medi_comb} }
\end{figure}

For the background data, the dependence of $C_1$  on the measured flux in the D aperture $\bar F = F_*+F_\mathrm{B}$ shows that the largest positive excess is observed near  $\bar F \approx 1$.  It is confirmed by the sub-sample medians, where each sub-sample is 1/10 of the full data set sorted in the ascending order. The medians are plotted in the top panel of  Fig.~\ref{fig:medi_comb}. Here, the combined $C_1$ parameter for the C and D channels is used. There is virtually no excess for faint fluxes, then it increases by $\sim 0.01$, and then it decreases again. The impact of the star light helps to explain this dependence. Low values of $\bar F = F_*+F_\mathrm{B}$ are encountered only if the stellar light is much less than   $F_\mathrm{B}$. For higher fluxes $F_*$, the measured points are shifted to a middle of the graph. The stellar flux impact on the $C_1$~(\ref{eq:8}) becomes inobservable for a high sky background $F_\mathrm{B}$.

The median stellar fluxes inducing the observed excesses are plotted in the bottom panel of Fig.~\ref{fig:medi_comb}. One can say that the flux $F_*$ increases while $F_* < \bar F$ and then saturates at  a constant level. The stellar flux fraction in the measured signal is significant and reaches of $0.4$ at maximum. Since this maximum is located not in the beginning, the minimal values of the measured sky background distribution are not distorted.

High-altitude clouds leading small ($\sim 10$  per cent) flux variations also affect the measured sky fluxes. The sky brightness usually considerably increases due to the light back-scattered by the Earth surface. For this reason,  we consider separately  the sky measurements carried out during photometric sky conditions. Their number is significantly less ($4\,566$) in spite of the fact that photometric conditions occur in $\sim 50$  per cent of the clear time, because  the sky background was rarely measured in stable conditions. The behaviour of the median  statistic $\bar C_1$ has not changed for the selected clear-sky measurements. Moreover, the observed excess became a little larger. 

Finally, we accept this estimate of stellar flux contribution (both faint stars and scattered light) in the measured sky background as reliable enough to be used for its correction. We use the approximation in the form 
\begin{equation}
\bar F_* = 0.3\,(1-\exp(-4\,\bar F^2))
\label{eq:appro}
\end{equation}
and subtract this correction $\bar F_*$ from the flux $\bar F$ measured in channel D.

\subsection{Night-sky brightness in the MASS photometric band}

The night-sky brightness is conventionally expressed in magnitudes per square arcsec at zenith \citep{Krisciunas1997PASP, Sanchez2007PASP} because a considerable amount of the sky radiation is produced inside the Earth atmosphere. Given the instrumental constant $R$, the following simple expression is used to obtain the night-sky magnitude:
\begin{equation}
m^\circ_\mathrm{B}\left(z\right) = R - 1.086\, \ln F_\mathrm{B}\left(z\right).
\label{eq:back_mag}
\end{equation}
The smoothed instrumental constant has been obtained in Section~\ref{sec:extinction} for the full duration of the measurement campaign, except the period  staring in January 2013 with device MD41, and interpolated onto sky measurement moments. As a result, $11\,580$ night-sky brightness magnitudes were obtained, $4\,227$  of those corresponding to the photometric conditions.

Eq.~(\ref{eq:back_mag}) provides the night-sky brightness $m^\circ_\mathrm{B}$ at the given sky point. One may use simple expressions from the papers \citep{Garstang1989PASP, Pedani2014NewA} to calculate the night-sky brightness at the zenith. Our measurements have been carried out at air masses $M\left(z\right) < 1.3$ (except a small fraction near the extinction standards),  the median air mass is $1.09$. The analysis shows that the correction to zenith is less than $0.02$~mag and it is neglected here.

For measurements in moonless period under photometric conditions, the sky magnitude $m^\circ_\mathrm{B}$ distribution quartiles are correspondingly $10.34$, $10.00$, and $9.32$~mag. These values confirm the preliminary estimate that stellar light may considerably impact the night-sky brightness measurements. To transform the sky brightness $m_\mathrm{B}$ to the units of magnitudes per square arcsec, the additional constant $11.50$~mag corresponding the aperture area in arcseconds should be added.

The empirical differential distribution of $m_\mathrm{B}$ is narrow enough (its width is less than $1$~mag) under moonless sky, i.e. when the Moon elevation is less than $-5^\circ$. The most probable night-sky brightness is about $21.8$~mag. The right bound of the distribution (the darkest sky) corresponds to $22.4$~mag. Comparison between the distribution described above and the overall data distribution (with any sky quality data) shows that the latter has much more estimates brighter than $21$~mag. This is explained by the additional light scattered by light clouds and fumulus. At the same time, some excess of dark sky due to additional extinction in clouds is observed.

The background measurements cover more than 5 years. Initially, the nearest light pollution sources were located on the Solar station, 800 meters away. The local light pollution has changed dramatically since autumn 2011 when observatory construction has begun in 40~m from the ASM. In the summer of 2012, a bright lamp was installed on the ASM tower to permanently floodlight the construction area.

\begin{figure}
\centering
\psfig{figure=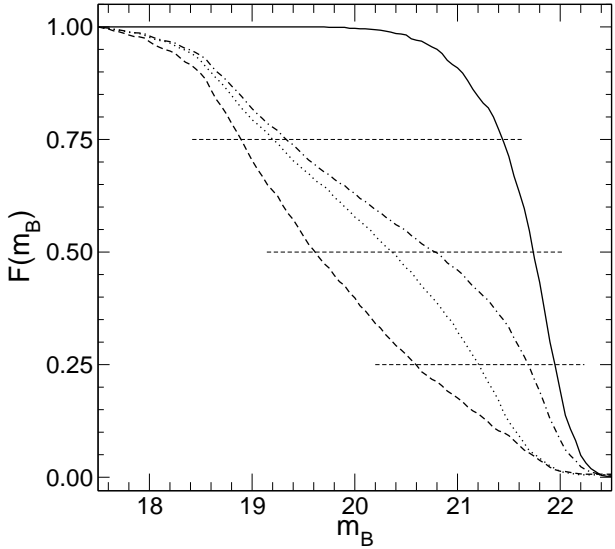,width=7.60cm,angle=-90}
\caption{Night-sky brightness complementary cumulative distributions for photometric conditions in moonless periods (solid line), in moonlight periods (dashed line), and for all data (dot-dashed). The dotted line shows the complete distribution without stellar light correction. The horizontal lines mark the quartile levels.
\label{fig:back_evol} }
\end{figure}

To check the local light pollution influence, we divided the moonless night-sky brightness measurements into five one-year seasons beginning from the 1-th of July. Per-season empirical cumulative distribution functions show that the first three seasons do not differ between themselves and the 25 per cent quartile is equal to $22.0$~mag.  The season 2011/12 demonstrates a brightening by $0.2$~mag. The sky brightness increased by another $0.5$~mag in the season 2012/13. These seasons were excluded from the final statistics. Of course, the light discipline has been recovered after the telescope installation.

This night-sky brightening is unlikely related to the Solar activity  maximum in 2013. The Solar activity impact has been widely discussed by many authors \citep{Krisciunas1997PASP, Patat2003AA}. However, let us note that the MASS spectral band (as well as $B$ band) are mostly free of the night-sky emission  lines.

Complementary cumulative distributions of the final samples are presented in Fig.~\ref{fig:back_evol} for different conditions. To estimate the impact of stellar light, the distributions of the sky brightness corrected with Eq.~(\ref{eq:appro}) and uncorrected  are shown. The  quartiles of these distributions are given in Table~\ref{tab:back_phot}. 

\begin{table}
\caption{Quartiles of the night-sky brightness distribution in the MASS photometric band, expressed in units of magnitudes per square arcsec. The column $N$ shows the sample size. \label{tab:back_phot}}
\centering
\begin{tabular}{lrrrr}
\hline
Sample             & 25 \% &  50 \%  &  75 \% & $N$ \\
\hline
Uncorrected             & 21.20    &   20.37  &  19.20 &  2439 \\
Corrected               & 21.69    &   20.78  &  19.33 &  2439 \\
Moon above horizon      & 20.59    &   19.62  &  18.89 &  1494 \\
Moon below horizon      & 21.95    &   21.74  &  21.44 &   945 \\
\hline
\end{tabular}
\end{table}

The inter-quartile range of the moonless night-sky brightness distribution  is $0.51$~mag or $1.6$ times. The most probable brightness is $21.85$~mag$\cdot$arcsec${}^{-2}$.

\subsection{ Night-sky brightness in the standard photometric system}

The night-sky brightness can be transformed  to the standard photometric spectral bands $B$, $V$, $R$, and $I$. We use the spectra of moonless night-time sky  obtained using FORS1 \citep{Patat2008AA} at ESO Paranal observatory in the 2000-s to calculate the sky colours. These data are freely available on the web\footnote{\url{http://www.eso.org/~fpatat/science/skybright/}} and consist of hundreds of spectra in absolute energy units. The zero point is found using the $\alpha$~Lyr spectrum from the paper \citep{Hayes1985}. The magnitudes of $\alpha$~Lyr are supposed to be $0.0$~mag for all bands except MASS, where it is equal to $0.05$~mag \citep{Voziakova2012AstL}. To obtain the $B-M$ and $M-V$ colours (where $M$ is the MASS band magnitude), one needs to integrate the night-sky spectra with corresponding reaction curves. The mean night-sky colours are given in Table~\ref{table:color}.

\begin{table}
\caption{Moonless night-sky colours. Values and standard deviations~(STD) are in units of magnitudes. $M$ is MASS band magnitude.
\label{table:color}}
\centering
\begin{tabular}{lrrrrr}
\hline
Colour         & $B-M$ & $M-V$ & $B-V$ & $V-R$ & $V-I$ \\
\hline
Value          & 0.37  & 0.66 & 1.03 & 0.77 & 2.10 \\
STD            & 0.04  & 0.09 & 0.12 & 0.13 & 0.20 \\
\hline
\end{tabular}
\end{table}

\begin{figure}
\centering
\psfig{figure=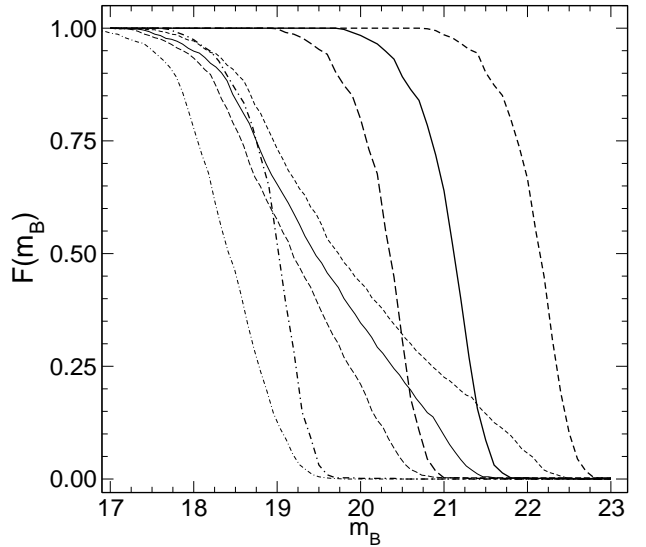,width=7.60cm,angle=-90}
\caption{ Complementary cumulative distribution of the night-sky brightness in different spectral bands: $B$ (dashed), $V$ (solid), $R$ (long-dashed) and $I$ (dot-dashed). The thick lines denote distributions for moonless periods, the thin lines designate the moonlight case.
\label{fig:ubvr_sky} }
\end{figure}

It is worth  noting that the calculated night-sky colours agree well with the previously published values given in the paper \citep{Patat2003AA}. Time-varying night-sky emission lines  affect the magnitudes in bands $V$, $R$, and $I$.  The moonless night-sky brightness estimates in the standard photometric bands, obtained using the calculated colours, are given in Table~\ref{tab:back_stan_phot}.

To obtain the night-sky colours in presence of the Moon, the numerical ESO model \citep{Jones2013AA} was used. The model calculates the synthetic sky spectrum for a given phase angle, Moon elevation, and Moon--target distance\footnote{\url{https://www.eso.org/observing/etc/skycalc/skycalc.htm}}. Using the synthetic spectra, we calculated the night-sky colours for the actual observing conditions and transformed our measurements to the standard photometric  bands. The results are presented in Fig.~\ref{fig:ubvr_sky} in the form of complementary cumulative distributions. The quartiles of these distributions are given in Table~\ref{tab:back_stan_phot}.

We remind that these statistics were calculated only on the data before autumn 2011, thus excluding the period of strong local light pollution. For the complete sample including the rejected data, the statistics change slowly, the quartiles  becoming brighter by $\sim 0.1$~mag in moonless time. Under moonlight, the changes are  less than a few $\sim 0.01$~mag.


\begin{table}
\caption{Quartiles of the night-sky brightness distribution in photometric bands $B$, $V$, $R$, and $I$. The values are in units of magnitude per square arcsec.  
\label{tab:back_stan_phot}}
\centering
\begin{tabular}{l|rrr|rrr}
\hline
Photometric  &  \multicolumn{3}{c}{Moon below horizon} &  \multicolumn{3}{c}{Moon above
horizon} \\
band       & 25 \%   &  50 \%   &   75 \% &  25 \% &  50 \% & 75 \% \\
\hline
$B$              & 22.31   &   22.10  &  21.84 & 20.87   &   19.73  &   18.96 \\
$V$              & 21.28   &   21.07  &  20.81 & 20.38   &   19.45  &   18.76  \\
$R$              & 20.51   &   20.30  &  20.04 & 19.85   &   19.18  &   18.57  \\
$I$              & 19.18   &   18.97  &  18.71 & 18.78   &   18.43  &   18.06  \\
\hline
\end{tabular}
\end{table}

\section{Conclusions}

The above results  illustrate that a slight modification of the optical turbulence measurement algorithm allows us to obtain additional information  from MASS/DIMM measurements. A possibility to study  photometric properties of the atmosphere with MASS/DIMM was pointed out by us a  long time ago \citep{2007bMNRAS}.

In order to determine the extinction in the MASS photometric band, the widely used method of pairs was adopted. Its ability to yield nearly instantaneous extinction estimates with minimum resources  was fully employed. This method requires knowledge of precise above-atmosphere instrumental magnitudes of the reference stars. At least a year-long cycle of measurements is needed to obtain a consistent solution of the respective system of photometric equations. Meanwhile,  the weather at Mt.~Shatdzhatmaz is such that a significant part of observing time occurs on partially clear nights, when the use of other methods of extinction estimates is not feasible.  

The transformation of the measured MASS band extinction coefficients into the standard $B$, $V$, $R$, and $I$ photometric  system leads to the following median extinction values: $0.28$, $0.17$, $0.13$, and $0.09$~mag, respectively.

It is worth to recall that a specific task to monitor the night sky brightness has not been initially planned  and, as result, 
the data presented in Section~\ref{sec:background} suffer from uncertainty and incompleteness. In particular, the unaccounted for  light scattering  in the telescope and instrument led to significant errors in the individual estimates of the sky brightness during moonless periods. Still, it was possible to derive  reasonable statistical estimates of the night sky background. Being transformed into the standard photometric system, they are $22.1$, $21.1$, $20.3$, and $19.0$~mag per square arcsec in the $B$, $V$, $R$, and $I$ bands, respectively.

The scattered light and star contamination are not important while measuring the sky background with moon above horizon. However,  building  a consistent model of the moon background at our site, similar to \citep{Jones2013AA}, is not straightforward because it contains many input parameters and requires additional measurements. 

Monitoring of the  photometric atmospheric properties will definitely be continued. The algorithm of the background measurement has already been improved. Measurements are now performed at 27 arcmin offset from the program star and in two positions. The extinction cycle now includes two low altitude stars in different parts of the sky.

At the same time, the emphasis of photometric measurements has now shifted to the real time use of the data for flexible planning of observations,  while achieving  statistical completeness has a lower priority. Special programs may acquire more time during optical turbulence monitoring and optimally will include a feedback from observations schedulers working for the telescopes at this site.

\section{Acknowledgements}

Authors are sincerely grateful to the SAI scientific community which welcomed our activity in astroclimatic research at Mt.~Shatdzhatmaz described in this and other papers. Especially we would like to thank A.\,Tokovinin for valuable remarks on the content and structure of the current article. A considerable help in this work was given by P.\,Kortunov, A.\,Belinski and, during the last two years, local observatory staff. As before, the support from the staff of the Pulkovo solar station which opened the way to start the current research at the site is acknowledged. 

\bibliography{references}
\bibliographystyle{mnras}
\label{lastpage}

\end{document}